\documentclass[showpacs,preprintnumbers,amsmath,amssymb]{revtex4}

\usepackage{epsfig}
\usepackage{graphicx}
\usepackage{dcolumn}
\usepackage{bm}

\begin{document}

\title{Redundant variables and Granger causality}
\author{
L. Angelini$^{1,2}$, M. de Tommaso$^{3}$, D. Marinazzo$^{4}$, L.
Nitti$^{1,5}$, M. Pellicoro$^{1,2}$, and S. Stramaglia$^{1,2}$}
\affiliation{$^1$ Istituto Nazionale di Fisica Nucleare, Sezione di
Bari, Italy\\} \affiliation{$^2$ Dipartimento di Fisica, University
of Bari, Italy\\}\affiliation{$^3$ Dipartimento di Scienze
Neurologiche e Psichiatriche, University of Bari, Italy \\}
  \affiliation{$^4$ Laboratoire de Neurophysique et Neurophysiologie, Universit\'e Paris Descartes, Paris, France\\}
   \affiliation{$^5$ Dipartimento di Biochimica Medica, Biologia Medica e Fisica Medica, University of Bari, Italy \\}
\date{\today}

\begin{abstract}
We discuss the use of multivariate Granger causality in presence of
redundant variables: the application of the standard analysis, in
this case, leads to under-estimation of causalities. Using the
un-normalized version of the causality index, we quantitatively
develop the notions of redundancy and synergy in the frame of
causality and propose two approaches to group redundant variables:
(i) for a given target, the remaining variables are grouped so as to
maximize the total causality and (ii) the whole set of variables is
partitioned to maximize the sum of the causalities between subsets.
We show the application to a real neurological experiment, aiming to
a deeper understanding of the physiological basis of  abnormal
neuronal oscillations in the migraine brain. The outcome by our
approach reveals the change in the informational pattern due to
repetitive transcranial magnetic stimulations.
\end{abstract}

\maketitle

Wiener \cite{wiener} and Granger \cite{granger} formalized the
notion that if the prediction of one time series could be improved
by incorporating the knowledge of past values of a second one, then
the latter is said to have a {\it causal} influence on the former.
Initially developed for econometric applications, Granger causality
has gained popularity also among physicists (see, e.g.,
\cite{chen,blinoska,smirnov,dingprl,lungarella}). A kernel method
for Granger causality, introduced in \cite{noiprl}, deals with the
nonlinear case by embedding data onto an Hilbert space, and
searching for linear relations in that space. Geweke \cite{geweke}
has generalized Granger causality to a multivariate fashion in order
to identify conditional Granger causality; as described in
\cite{noipre}, multivariate causality may be used to infer the
structure of dynamical networks \cite{dn} from data.

Granger causality is connected to the information flow between
variables \cite{hla}. Another important notion in information theory
is the redundancy in a group of variables, formalized in
\cite{palus} as a generalization of the mutual information. A
formalism to recognize redundant and synergetic variables in
neuronal ensembles has been proposed in \cite{sch} and generalized
in \cite{bettencourt}; the information theoretic treatments of
groups of correlated degrees of freedom can reveal their functional
roles in complex systems.

The purpose of this work is to show that the presence of redundant
variables influences the performance by multivariate Granger
causality and to propose a novel approach to exploit redundancy so
as to identify functional patterns in data. In the following we
provide a quantitative definition to recognize redundancy and
synergy in the frame of causality and show that the maximization of
the total causality is connected to the detection of groups of
redundant variables.

Let us consider $n$ time series $\{x_\alpha (t)\}_{\alpha
=1,\ldots,n}$ \cite{nota}; the state vectors are denoted
$$X_\alpha (t)= \left(x_\alpha (t-m),\ldots,x_\alpha (t-1)\right),$$
$m$ being the window length (the choice of $m$ can be done using the
standard cross-validation scheme). Let $\epsilon \left(x_\alpha
|{\bf X}\right)$ be the mean squared error prediction of $x_\alpha$
on the basis of all the vectors ${\bf X}$ (corresponding to linear
regression or non linear regression by the kernel approach described
in \cite{noiprl}). The multivariate Granger causality index $\delta
(\beta \to \alpha )$ is defined as follows: consider the prediction
of $x_\alpha$ on the basis of all the variables but $X_\beta$ and
the prediction of $x_\alpha$ using all the variables, then the
causality is  the (normalized) variation of the error in the two
conditions, i.e.
\begin{equation}\label{delta}
\delta (\beta \to \alpha )={\epsilon \left(x_\alpha |{\bf
X}\setminus X_\beta\right)-\epsilon \left(x_\alpha |{\bf
X}\right)\over \epsilon \left(x_\alpha |{\bf X}\setminus
X_\beta\right)}.
\end{equation}
Here we use the selection of significative eigenvalues described in
\cite{noiprl}  to address the problem of over-fitting in
(\ref{delta}).

The straightforward generalization of Granger causality for sets of
variables is
\begin{equation}\label{sets}
\delta (B \to A)={\Large\sum_{\alpha \in A}}{\epsilon \left(x_\alpha
|{\bf X}\setminus B\right)-\epsilon \left(x_\alpha |{\bf
X}\right)\over \epsilon \left(x_\alpha |{\bf X}\setminus B\right)},
\end{equation}
where $A$ and $B$ are two disjoint subsets of $\{1,\ldots,n\}$, and
${\bf X}\setminus B$ means the set of all variables except for those
$X_\beta$ with $\beta\in B$.

 On the other hand, the un-normalized version of it,
i.e.
\begin{equation}\label{unnorm}
\delta^u (B \to A)=\sum_{\alpha \in A}\{\epsilon \left(x_\alpha
|{\bf X}\setminus B\right)-\epsilon \left(x_\alpha |{\bf
X}\right)\},
\end{equation}
can be easily be shown to satisfy the following interesting
property: if $\{X_\beta\}_{\beta\in B}$ are statistically
independent and their contributions in the model for A are additive,
then
\begin{equation}\label{deltaset}
\delta^u (B \to A)=\sum_{\beta \in B} \delta^u (\beta \to A).
\end{equation}

In order to identify the informational character of a set of
variables $B$, concerning the causal relationship $B \to A$, we
remind that, in general, synergy occurs if $B$ contributes to $A$
with more information than the sum of all its variables, whilst
redundancy corresponds to situations with the same information being
shared by the variables in $B$. Following
\cite{palus,sch,bettencourt}, we make quantitative these notions and
define the variables in $B$ {\it redundant} if $\delta^u (B \to A) >
\sum_{\beta \in B} \delta^u (\beta \to A)$, and {\it synergetic} if
$\delta^u (B \to A) < \sum_{\beta \in B} \delta^u (\beta \to A)$. In
order to justify these definitions, firstly we observe that the case
of independent variables (and additive contributions) does not fall
in the redundancy case neither in the synergetic case, due to
(\ref{deltaset}), as it should be. Moreover, we describe the
following example for two variables $X_1$ and $X_2$. If $X_1$ and
$X_2$ are redundant, then removing $X_1$ from the input variables of
the regression model does not have a great effect, as $X_2$ provides
the same information as $X_1$; this implies that $\delta^u (X_1 \to
A)$ is nearly zero. The same reasoning holds for $X_2$, hence we
expect that $\delta^u (\{X_1,X_2\} \to A) > \delta^u (X_1 \to A)+
\delta^u (X_2 \to A)$. Conversely, let us suppose that $X_1$ and
$X_2$ are synergetic, i.e. they provide some information about $A$
only when both the variables are used in the regression model; in
this case $\delta^u (\{X_1,X_2\}\to A)$, $ \delta^u (X_1 \to A)$ and
$\delta^u (X_2 \to A)$ are almost equal and therefore $\delta^u
(\{X_1,X_2\} \to A) < \delta^u (X_1 \to A)+ \delta^u (X_2 \to A)$.

Two analytically tractable cases are now reported as examples.
Consider two stationary and Gaussian time series $x(t)$ and $y(t)$
with $\langle x^2(t)\rangle=\langle y^2(t)\rangle=1$ and $\langle
x(t)y(t)\rangle={\cal C} $; they correspond, e.g., to the asymptotic
regime of the autoregressive system
\begin{eqnarray}
\begin{array}{ll}
x_{t+1}&=a x_t+b y_t +\sigma \xi^{(1)}_{t+1}\\
y_{t+1}&=b x_t+a y_t +\sigma \xi^{(2)}_{t+1},
\end{array}
\label{map}
\end{eqnarray}
where $\xi$ are i.i.d. unit variance Gaussian variables, ${\cal
C}=2ab/(1-a^2-b^2)$ and $\sigma^2=1-a^2-b^2-2ab{\cal C}$.
Considering the time series
$z_{t+1}=A\left(x_t+y_t\right)+\sigma^\prime \xi^{(3)}_{t+1}$ with
$\sigma^\prime=\sqrt{1-2A^2(1+{\cal C})}$, we obtain for $m=1$:
\begin{equation}\label{z}
\delta^u (\{x,y\} \to z)-\delta^u (x \to z)-\delta^u (y \to
z)=A^2({\cal C}+{\cal C}^2).
\end{equation}
Hence $x$ and $y$ are redundant (synergetic) for $z$ if ${\cal C}$
is positive (negative). Turning to consider $w_{t+1}=B\; x_t \cdot
y_t+\sigma^{\prime \prime} \xi^{(4)}_{t+1}$ with $\sigma^{\prime
\prime}=\sqrt{1-B^2(1+2{\cal C})^2}$, and using the polynomial
kernel with $p=2$, we have
\begin{equation}\label{z}
\delta^u (\{x,y\} \to z)-\delta^u (x \to z)-\delta^u (y \to
z)=B^2(4{\cal C}^2-1);
\end{equation}
$x$ and $y$ are synergetic (redundant) for $w$ if $|{\cal C}| < 0.5$
($|{\cal C}| > 0.5$).

\begin{figure}[ht!]
\begin{center}
\epsfig{file=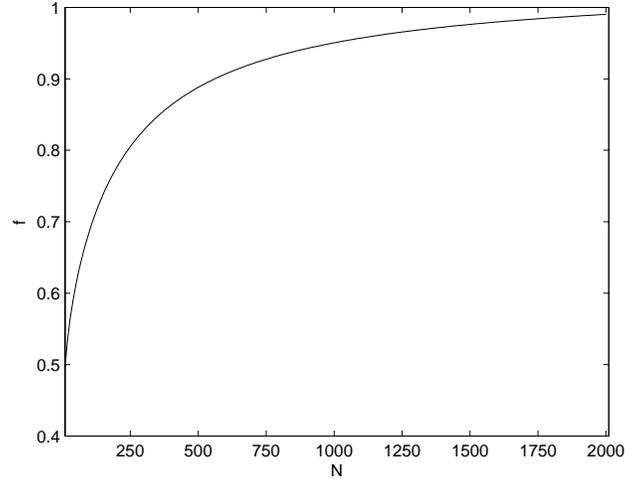,height=7.cm}
\end{center}
\caption{{\small The fraction $f$ of times that the $x$ and $y$ are
recognized as redundant for the variable $z$ (see the text), versus
the number of samples $N$. $f$ is evaluated over $10^6$ repetitions.
\label{fig1}}}
\end{figure}

The presence of redundant variables leads to under-estimation of
their causality when the standard multivariate approach is applied
(this is not the case for synergetic variables). Redundant variables
should be grouped to get a reliable measure of causality, and to
characterize interactions in a more compact way. As it is clear from
the discussion above, grouping redundant variables is connected to
maximization of the un-normalized causality index (\ref{unnorm})
and, in the general setting, can be made as follows. For a given
target $\alpha_0$, we call $B$ the set of the remaining $n-1$
variables. The partition $\{A_\ell\}$ of $B$, maximizing the total
causality
$$\Delta =\sum_\ell \delta^u (A_\ell \to x_{\alpha_0}),$$
consists of groups of redundant variables. Concerning the problem of
finite sample size, we consider $N$ samples from eqs. (\ref{map}),
with $a=0.5$ and $b=0.4$, and estimate casualities on these data. In
figure (\ref{fig1}) we depict, as a function of $N$, the fraction
$f$ of times that the $x$ and $y$ are recognized as redundant for
the variable $z$ (with $A=0.4$); a large amount of data is needed to
assess significative causality and so to discover redundancy. The
present approach can thus be used only in applications such that a
large number of samples is available.

\begin{figure}[ht!]
\begin{center}
\epsfig{file=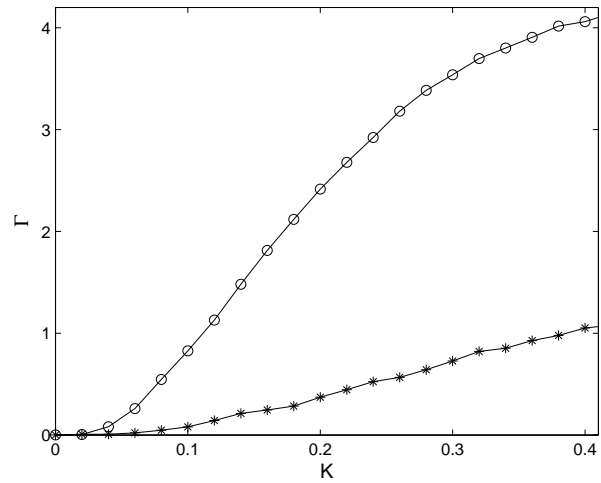,height=7.cm}
\end{center}
\caption{{\small Concerning the system of nine oscillators described
in the next, we depict the sum of the  causalities between every
pair of subsets $\Gamma$ (see the text) corresponding to the
partitions $\{1,2,3\}\{4,5,6\}\{7,8,9\}$ (empty circles) and
$\{1\}\{2\}\{3\}\{5\}\{6\}\{7\}\{8\}\{9\}$ (stars). Causalities are
estimated over $5000$ samples for each value of $K$.\label{fig2}}}
\end{figure}

Another example consists of a system of nine oscillators evolving
according to noisy Kuramoto's equations \cite{kuramoto}:
\begin{equation}\label{kuramoto}
\dot{\theta_i}=\omega_i + K \sum_{j=1}^9 \;
sin\left(\theta_j-\theta_i \right) + \xi_i(t);
\end{equation}
We consider three groups of oscillators, each made of three
oscillators with the same natural frequency, respectively $\omega =
1, 2, 4$; the noise strength is $0.01$. Using the approach for
circular variables, described in \cite{noipla}, we find that the
partition $\{A_\ell\}$ of the nine oscillators, maximizing the sum
of the  causalities between every pair of subsets
$$\Gamma= \sum_\ell \sum_{\ell^\prime \ne \ell} \delta^u (A_\ell \to
A_{\ell^\prime}),$$ is $\{1,2,3\}\{4,5,6\}\{7,8,9\}$, corresponding
to oscillators with the same natural frequency belonging to the same
subset. In figure (\ref{fig2}) we  depict the optimal $\Gamma$ and
the value of $\Gamma$ corresponding to the partition where each
oscillator constitutes a set, versus the coupling $K$. It is clear
that  the maximization of $\Gamma$  reveals the structure of the
system in this example.

\begin{figure}[ht!]
\begin{center}
\epsfig{file=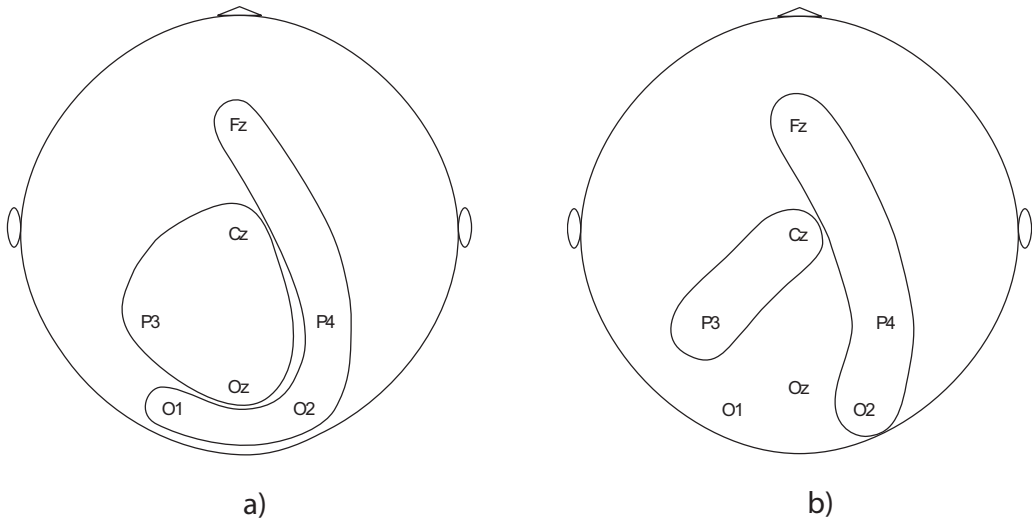,height=7.cm}
\end{center}
\caption{{\small  The partitions of electrodes maximizing $\Gamma$
(see the text). Left: the optimal partition for Basal and Sham
conditions. Right: the optimal partition in presence of
TMS.\label{fig3}}}
\end{figure}

Now we turn to consider a real application, i.e. EEG data from
nineteen subjects suffering from migraine, under steady state flash
stimuli (9 Hz)  and repetitive transcranial magnetic stimulation
(rTMS), a noninvasive method to excite neurons in the brain
\cite{lanc}. Migraine is a complex disorder of neurovascular origin
whose pathophysiological basis is largely unknown. An altered
cortical excitability may activate the trigemino-vascular system,
but the question about a basal hypo or hyper cortical excitability
is actually a matter of debate \cite{coppola}. In a previous work
\cite{prlss} anomalous cortical synchronization in migraneurs under
flash stimuli has been reported. A better understanding of migraine
pathophysiology may improve its therapeutical approach: in this
view, studies employing neurophysiological techniques, possibly
supported by advanced methods of quantitative analysis, may give an
aid to the knowledge of migraine pathophysiology \cite{valeriani}.
An important feature of migraine brain, is the tendency to
hypersynchronization of alpha rhythms, which is influenced by
anti-epileptic drugs \cite{detommaso}. rTMS  induces a cortical
modulation that lasts beyond the time of stimulation \cite{fregni}:
its effects depend on the frequency of stimulations. In order to
understand the physiological basis of  abnormal neuronal
oscillations in migraine brain, we apply 1 Hz rTMS over the
occipital cortex, before performing repetitive flash stimulation.
The records are 12 seconds long, sampled at 256 Hz: this EEG
duration is representative of the pattern of brain responsiveness to
light stimuli, as previously shown \cite{prlss}. The signals are
measured on seven channels (Fz,P3,P4,Cz,O1,Oz,O2) in three
conditions basal (only flash stimuli) sham (placebo, i.e. flash
stimuli and a fake magnetic stimulator) and rTMS (flash stimuli and
magnetic stimulations) . As in the example above, for each target
channel we exhaustively search for the partition of the remaining
six channels which leads to the highest total causality $\Delta$
(averaged over the nineteen patients). In basal and sham conditions,
we find that, for each target channel, the optimal partition is
always a single set containing all the six remaining channels, in
other words all the channels are redundant in these conditions. In
presence of rTMS the causality pattern becomes more complex, and not
all sets of variables are redundant w.r.t. the prediction of the
others. All the six remaining channels are redundant for targets
Fz,P4,O1,Oz; for the other channels the best partitions are
\begin{eqnarray}
\begin{array}{rll}
\{ Fz,P4,O1,Oz,O2\}\{P3\}& \to &Cz\\
\{ Fz,P4,O1,Oz,O2\}\{Cz\} &\to &P3\\
\{ Fz,P4,O1,Oz\}\{Cz P3\}& \to &O2.
\end{array}
\label{map3}
\end{eqnarray}
These relations suggest the presence of a new source of information,
due to magnetic stimulations, corresponding to Cz and P3 channels.
We also search for the partition of the seven channel maximizing the
total causality between groups ($\Gamma$), averaged over the
patients. We find that the best partition is $\{ Fz,P4,O1,O2\}\{Cz
P3 Oz\}$ for basal and sham conditions. For the TMS condition,
instead, the best partition is $\{ Fz,P4,Oz\}\{Cz P3\}\{O1\}\{O2\}$;
this result is consistent with the previous analysis as the channels
Cz and P3 are grouped, see figure (\ref{fig3}).

The change of the informational pattern, induced by occipital cortex
inhibition, may confirm that neuronal oscillations are related to
the state cortical excitability. Presently, we have no explanation
about the significance of the specific Cz-P3 group related to  rTMS
effect, but we can assert that  oscillations in migraine brain vary
as a function of cortical excitability. The reliability of this
pattern in migraine needs to be matched with a control group, so as
to better understand the peculiar reactivity of migraine brain and
to find the optimal way to influence it.
Some remarks are in order.  Averaging over patients is mandatory to
reduce the effects due to the variability among subjects. Our
results are obtained using the linear kernel and $m=1$, but the same
partitions are obtained using the quadratic kernel and $m=2$
(application of cross-validation, on these data, suggests a low
value of the order m; therefore we restrict our analysis  to
$m=1,2$). We find, in this real application, that the optimal
partition maximizing the total causality is unique in all cases. It
may happen, in other instances, that several partitions have the
same total causality: in those cases prior information should be
used to select one of the degenerate partitions.

Summarizing, in this work we have quantitatively developed the
notions of redundancy and synergy in the frame of causality. We have
 proposed to generalize the standard multivariate Granger method in presence
of redundant variables, by using the causality index without
normalization, and analyzing the system as follows: (i) for a given
target, the remaining variables are grouped so as to maximize the
total causality and (ii) the whole set of variables is partitioned
to maximize the sum of the causalities between groups. Analyzing
real data from a neurophysiological experiments, the proposed
approach was able to detect the informational pattern induced by
magnetic stimulations.


\begin{thebibliography}{99}
\bibitem{nota} After a linear transformation, we may assume all the
time series to have zero mean and unit variance.
\end{thebibliography}
\end{document}